\documentclass[aps,twocolumn,amssymb,amsmath,showpacs]{revtex4}

\usepackage{graphicx}% Include figure files

\def \square {\hbox{$\sqcup\!\!\!\!\sqcap$}}

\newcommand{\be}{\begin{equation}}
\newcommand{\ee}{\end{equation}}
\newcommand{\bea}{\begin{eqnarray}}
\newcommand{\eea}{\end{eqnarray}}

\begin{document}

\title{Non-local density correlations as signature of Hawking radiation\\ from acoustic black holes}
%\maketitle

\author{Roberto Balbinot$^a$, Alessandro Fabbri$^b$, Serena
Fagnocchi$^{c,a}$, Alessio Recati$^{d}$, and Iacopo Carusotto$^{d}$}
%\altaffiliation{Email addresses: balbinot@bo.infn.it, afabbri@ific.uv.es,\\ fagnocchi@bo.infn.it}
\affiliation{{\it a) } Dipartimento di Fisica dell'Universit\`a di Bologna and INFN sezione di Bologna,  Via Irnerio 46, 40126 Bologna, Italy\\
{\it b)} Departamento de Fisica Teorica and IFIC, Universidad de
Valencia-CSIC,  C. Dr. Moliner 50, 46100 Burjassot, Spain\\
{\it c)} Centro Studi e Ricerche {\it "Enrico Fermi"}, Compendio Viminale, 00184 Roma, Italy\\
{\it d)} CNR-INFM BEC Center and Dipartimento di Fisica, Universit\`a di Trento, via Sommarive 14,
I-38050 Povo, Trento, Italy }
%E-mail: balbinot@bo.infn.it, afabbri@ific.uv.es, fagnocchi@bo.infn.it}

\begin{abstract}
We have used the analogy between gravitational systems and
non-homogeneous fluid flows to calculate the density-density
correlation function of an atomic Bose-Einstein condensate in the
presence of an acoustic black hole.
The emission of correlated pairs of phonons by Hawking-like process results
into a peculiar long-range density correlation.
Quantitative estimations of the effect are provided for realistic experimental
 configurations.
\end{abstract}

% 05.30.Jp,

\maketitle

Hawking's prediction of black holes evaporation is generally regarded as
a milestone of modern theoretical physics.
Combining Einstein's General Relativity and Quantum Mechanics, Hawking
was able to show that black holes are not "black", but emit thermal
radiation at a temperature inversely proportional to their mass
\cite{hawking}.
This quantum mechanical process is triggered by the formation of a
horizon and proceeds via the conversion of vacuum fluctuations into
on-shell particles. Unfortunately so far there is no experimental support for
this amazing theoretical prediction. The emission temperature (Hawking
temperature) for a solar mass black hole is expected to be of the order
of $10^{-8}$ K, far below the $3$ K cosmic microwave background.
Nor evidence has been found so far of a X-ray background from a hypothetical
primordial population of low mass black holes
($\sim 10^{10} $ Kg) in the final stages of their evaporation
\cite{carr}. Expectations to directly observe Hawking radiation from
mini-black holes formed in colliders like LHC or next generation ones,
are based on models where the quantum gravity scale (Planck scale:
$10^{19}$ GeV) is lowered down to the TeV scale by the presence  of
extra-dimensions \cite{TeVgravity}.
It is perhaps fair to say that the prospects to have a direct
experimental detection of Hawking radiation from black holes in the near
future are not very optimistic.

In a remarkable work  Unruh \cite{unruh81} showed that Hawking radiation
is not peculiar to gravity, but is rather a purely kinematic effect of
quantum field theory which only depends on field propagation on a black
hole-type curved space-time background. This opens the concrete possibility to study the Hawking
radiation process in completely different physical systems.
As an example, the propagation of sound waves in Eulerian fluids
can be described in terms of the same equation describing a massless
scalar field on a curved spacetime characterized by an \emph{acoustic
metric} $G_{\mu\nu}$ which is function of the background flow: the
curvature of the acoustic geometry is induced by the inhomogeneity of
the fluid flow, while flat minkowskian spacetime is recovered in the
case of a homogeneous system.
In particular, an {\em acoustic black hole} (or {\em dumb hole})
configuration is obtained whenever a subsonic flow turns supersonic:
sound waves in the supersonic region are in fact dragged away by the
flow and can not propagate back towards the {\em acoustic horizon}
separating the supersonic and subsonic regions. Upon quantization,
Hawking radiation is expected to appear as a flux of thermal phonons
emitted from the horizon at a temperature proportional to its
surface gravity.
Even though a substantial effort has been spent on a variety of
{\em analog models}~\cite{analogy_book}, e.g. superfluid
Helium~\cite{3He}, phonons in atomic Bose-Einstein
condensates~\cite{bec-analogy}, degenerate Fermi gases~\cite{giova}, 
slow light in moving media~\cite{slowlight}, or travelling refractive
index interfaces in nonlinear media~\cite{fiber}, the weakness of Hawking
radiation has so far prevented experimental verification of these
predictions.

In the present Letter we propose a novel route to detect the emission
by an acoustic black hole in a flowing Bose-Einstein condensate.
Differently from the case of astrophysical black holes, both the
external and the internal region of the acoustic black hole are in fact
accessible to experiment: the quantum-optical correlations between the
Hawking radiation emitted inside and outside the acoustic black hole are
responsible for a significant long-range correlation between the density
fluctuations at points respectively inside and outside the acoustic
black hole.
This unique signature can be exploited to isolate
Hawking radiation from the background of competing processes and
experimental noise.

\begin{figure}[t]
\begin{center}
\includegraphics[ width=80mm , height=48 mm ,clip ]{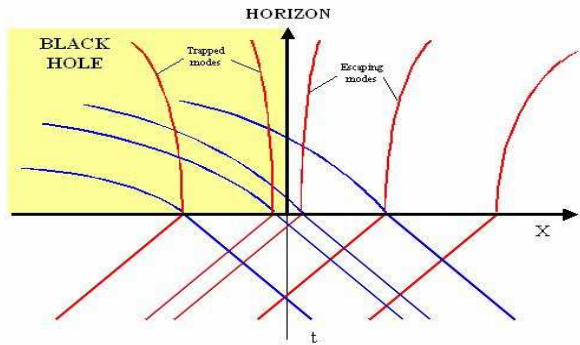}
\caption{Space-time diagram describing the ingoing (blue lines) and
 outgoing (red) modes. Before the black hole formation ($t<0$) modes
 propagate along straight lines. When the horizon (in $x=0$) is formed
 propagation gets significantly distorted and the outgoing modes inside
 the black hole remain trapped.} 
\label{shock}
\end{center}
\end{figure}

We start by briefly recalling those features of Hawking radiation
that we shall use in what follows. 
Consider the quantum field associated to Hawking radiation to be in its
vacuum state at early times before the horizon formation (let us call it
$|in\rangle$).
In the Heisenberg picture, the state of the field at late times will be
still $|in\rangle$, but because of the horizon formation the
$|in\rangle$ state does not correspond to the late time vacuum state
$|out\rangle$ of the field. A basis for the $out$
modes (see Fig.\ref{shock}) is composed by ingoing, i.e.,
propagating downstream in the free-falling frame, modes and outgoing, i.e., modes propagating upstream, modes.
Let $|0_{ing}\rangle$ and $|0_{outg}\rangle$ be the vacuum states of
respectively the ingoing and outgoing modes.
Among the outgoing modes, we must distinguish between the ones which
propagate outside the horizon with a sound velocity faster than the flow
velocity and are therefore able to reach the asymptotic region far from
the horizon, and the ones that are trapped inside the horizon and are
dragged inwards by the flow which is here faster than the speed of
sound.

Hawking radiation corresponds to particles creation in the outgoing
sector only and, limiting to this latter, 
the late-time decomposition for the $|in\rangle$ vacuum in terms of
these escaping and trapped modes has the form of a two-mode squeezed
vacuum state~\cite{librosandro}:
\be
|in\rangle\propto\, \exp\left(\frac{1}{\hbar}
\sum_\omega e^{-\frac{\hbar \omega}{2\kappa_B T_H}}
a_\omega^{(esc)\,\dagger}a_\omega^{(tr)\,\dagger}
\right)
|0_{outg}\rangle,
\label{squeeze}
\ee
where $a_\omega^{(esc,tr)\,\dagger}$ are creation operators for
respectively the outgoing escaping and trapped modes, $T_H$ is the
Hawking temperature and $\kappa_B $ the Boltzmann constant.
Thus, the Hawking process consists of the creation of correlated pairs
of outgoing quanta triggered by the horizon formation~\cite{CW,parentani}:
one member is emitted into an escaping mode, and corresponds to what is
usually called Hawking particle; the other member, the so-called
\emph{partner}, is emitted into a trapped mode inside the horizon. Of
course, this latter particle is definitively lost and can not be
detected in the case of astrophysical black holes.

Let us now move to an analog model of black hole based on a flowing
atomic Bose-Einstein condensate (BEC).
As we are considering small fluctuations around a stationary and fully
condensed state, we write the Bose field operator as
$\hat\Psi=e^{i\hat \theta}\,\sqrt{\hat n}$ in terms of the number
density $\hat n$ and the phase $\hat \theta$~\cite{stringari} and we  
expand the density $\hat n=n+\hat n_1$ and phase  
$\hat \theta=\theta+\hat \theta_1$ operators around the classical
background values $n$ and $\theta$ fixed by the mean-field Gross-Pitaevskii
equation.

We limit our attention to the so-called hydrodynamic limit, where perturbations are considered with wavelengths much longer than the
healing length $\xi=\hbar/\sqrt{mgn}$ ($m$ is the atomic mass and $g$
the atom-atom nonlinear interaction constant in the Gross-Pitaevskii
equation~\cite{stringari}). The linearized equation of motion for the
phase $\hat \theta_1$ field is then formally equivalent to a curved spacetime
field equation for a massless scalar field~\cite{bec-analogy}
\be
\label{meq}
\square \hat{\theta}_1=\frac{1}{\sqrt{-G}}\partial_\mu
(\sqrt{-G}\,G^{\mu\nu}\partial_\nu)\,\hat{\theta}_1=0\, ,
\ee
where $\square$  is the curved d'Alembertian for the acoustic metric
$G_{\mu\nu}$ of line element
\be
\label{mac}
ds^2=G_{\mu\nu}dx^\mu dx^\nu=\frac{n}{mc}[-c^2 dt^2+(d\vec x-\vec v
dt)\cdot(d\vec x-\vec v dt)]\, ,
\ee
and $G$ is the metric determinant, $\vec v=\hbar{\vec \nabla} \theta/m$ is the local flow velocity and $c=\sqrt{gn/m}$ is the local sound speed.
As the density operator $\hat n_1$ is algebraically related to $\hat
\theta_1$ by
%\be \label{n1}
$\hat n_1=-\frac{\hbar}{g}\,\left(\partial_t\hat
\theta_1+\frac{\hbar}{m}\,
{\vec\nabla} \theta
\cdot \vec \nabla \hat \theta_1
\right),$
the one-time density-density correlation function
\be
G_2(x,x')=\langle \hat n(x)\hat n(x')\rangle-\langle \hat n(x)\rangle
\langle \hat n(x')\rangle
\label{g2_a}
\ee
can be simply expressed in term of the two-points function for the field
$\hat \theta_1$:
\be
\label{g2}
G_2(x,x')=\frac{\hbar^2}{g(x,t)\,g(x',t)}\lim_{t'\rightarrow t}
\mathcal{D} \langle\hat \theta_1( x,t)\hat \theta_1(
x',t')\rangle\, ,
\ee
the operator $\mathcal{D}$ being defined as
$\mathcal{D}=\left[\partial_{t}\partial_{t'}+
v(\vec x)\vec \nabla_{\vec x} \partial_{t'}+v(\vec x') \partial_{t} \vec
\nabla_{\vec x'}+v(\vec x)v( \vec x') \vec \nabla_{\vec x}\cdot \vec
\nabla_{\vec x'}\right]$.

To obtain workable analytical expressions we restrict
our attention to the simplest case of a one-dimensional condensate,
whose transverse size $\ell_\perp$ is assumed to be much smaller than
the healing length $\xi$. Configurations of this kind can be realized
in the lab by outcoupling atoms from a mother condensate in to an atomic fiber to form a so-called
{\em atom laser} beam~\cite{atomlaser,atomlaser2,pavloff}.
Performing a dimensional reduction along the transverse
direction~\cite{review}, the 
two-point function for the field $\hat \theta_1$ can be approximated as:
\be
\label{thetasquared}
\langle \hat \theta_1(x,t)\hat\theta_1(x',t')\rangle
\simeq
-\frac{1}{4\pi \, \sqrt{C(x,t)C(x',t')}}\ln[\Delta x^- \Delta x^+]
\ee
where
$C=n_{1D}\xi$ is the conformal factor for the metric
(\ref{mac}), with $n_{1D}=n\ell_\perp^2$~\cite{pavloff}, and 
$x^\pm = t\pm \int \frac{dx}{(c\mp v)}$ are
light (sound) cone coordinates.
The expectation value entering the two-point function
(\ref{thetasquared}) is to be taken in the
vacuum state of the positive frequency modes with respect to the
$x^{\pm}$ coordinates~\cite{c-QFT,librosandro}.  
After dimensional reduction, an effective potential term appears
in the $1+1$ dimensional wave equation (\ref{meq}), which is responsible
for backscattering of the modes. In what follows we neglect such a term, to obtain a conformally invariant $1+1$ dimensional theory that 
can be handled in a fully analytical way~\cite{awp}.
This is expected to produce only slight quantitative overestimations of
the final result \cite{V}.

The consistency of this approach can be validated on the simplest case of a
spatially homogeneous one-dimensional BEC of density $n_{1D}$ moving at
a constant and spatially uniform speed $v$. In this case the
light cone coordinates associated to the vacuum are simply
$x^\pm=t\pm \frac{x}{c\mp v}$. Using (\ref{g2}), this leads to the result
%\be
${G^{1D}_2(x,x')}=-n_{1D}\xi/[2\pi(x-x')^2]$
%\ee 
which fully agrees with standard BEC theory~\cite{castin} in the
long-distance $|x-x'|\gg \xi$ limit where the hydrodynamic approximation is valid.

\begin{figure}[t]
\begin{center}
\includegraphics[ width=85mm , height=50mm,clip]{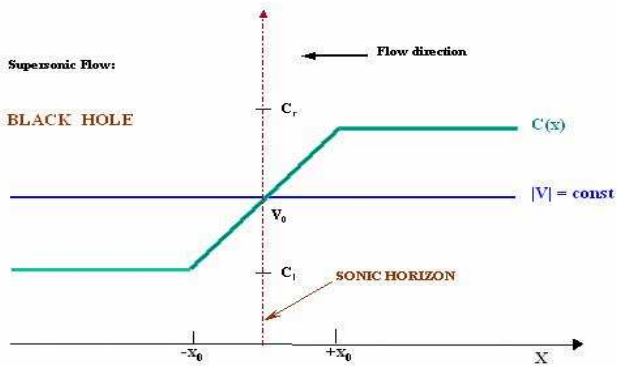}
\caption{Sketch of the proposed set-up: the density $n$ and the flow
 velocity $v$ are kept constant, while the sound velocity $c$ is
 modified around $t=0$ according to the picture. The sonic horizon is at $x=0$.  }
\label{fig1}
\end{center}
\end{figure}

We now turn to the most interesting case of an acoustic black hole.
Consider a one-dimensional BEC of uniform density $n_{1D}$ which is
flowing with constant speed $v$ along the $-\hat{x}$ direction.
For $t<0$ the sound velocity $c(x,t)$ is constant in space and equal to
$c(x,t)=c_r>v$.
At $t=0$, the sound velocity in the $x<0$ region is rapidly switched
to a lower $c_l<v$ value in order to create a horizon that separates a
region of sub-sonic $v<c_r$ flow for $x>0$ from a super-sonic $v>c_l$
one for $x<0$.
In the crossover region $x\in[-x_0,x_0]$, the sound velocity is assumed
to vary linearly in $x$. In order for the hydrodynamic approximation to be valid $x_0$ has to be larger than $\xi$. The sonic horizon (i.e. the locus where
$c=v$) lies at $x=0$ (see Fig. 2).

In practice, such a modulation can be obtained by spatially modulating
the transverse confinement of the waveguide~\cite{pavloff,Olshanii}
and/or the nonlinear
interaction constant $g(x)$ by means of a spatially varying
magnetic
field in the vicinity of a Feshbach resonance~\cite{stringari}.
In order to minimize competing processes such as back-scattering of
condensate atoms, Landau-\u Cerenkov phonon emission~\cite{Cerenk}, and
soliton shedding~\cite{pavloff} from the horizon region, the change in
$g(x)$ has to be compensated by a corresponding change in the external
potential $V(x)$ so to keep the local potential $n g(x)+ V(x)$ constant.

While the ingoing modes remain positive frequency modes with respect to the
$x^+$ light cone coordinate even after the
formation of the black hole, the outgoing modes are no longer positive
frequency with respect to the $x^-$ light cone coordinate.
Indeed they emerge as positive frequency with respect to the
generalized coordinate  $\tilde x^-=\pm
\frac{e^{-kx^-}}{k}$~\cite{c-QFT,review}, where the surface gravity at
the horizon (located at $x^-=\infty$ or $\tilde x^-=0$) is defined
as
\be
k=\frac{1}{2v}\frac{d}{dx}(c^2-v^2)|_H=\left(\frac{dc}{dx}\right)_H
\ee
and fully determines the Hawking temperature
$T_H=\hbar k /(2\pi \kappa_B)$ \cite{c-QFT}.
The $\pm$ signs in the definition of $\tilde x^-$ refer to
outgoing modes which are respectively trapped inside the horizon ($x<0$)
or able to escape towards the asymptotic subsonic region ($x>0$).

The correlation function $G^{1D}_2(x,x')$ can then be derived from the general
formulas Eqs.(\ref{g2}, \ref{thetasquared}) using
$\ln (\Delta\tilde x^- \Delta x^+)$
as the master function for the $|in\rangle$ state.
A compact formula can then be obtained for the most interesting case
when the $x,x'$ points are located on opposite sides with respect to
the horizon and lie well outside the modulation region i.e.  $x>x_0$
and $x'<-x_0$:
\begin{multline}
\label{corr}
\frac{G^{1D}_2(x,x')}{n_{1D}^2}\simeq -\frac{k^2\,\xi_l \xi_r}{16\pi\, c_l
 c_r}\,\frac{1}{\sqrt{(n_{1D}\xi_r)(n_{1D}\xi_l)}}\times \\
 \frac{c_r\,c_l}{(c_r-v)(v-c_l)}\, 
\frac{1}{\cosh^{2}\left[\frac{k}{2}\left(\frac{x}{c_r-v}+\frac{x'}{v-c_l}\right)
\right]}
% \\ +O\left[\frac{1}{|x-x'|^2}\right]
 %G_2(x,x')=-\frac{\hbar^2}{16\pi g_{l}g_{r}}\frac{1}{An\sqrt{\xi_{l}\xi_{r}}}
%\frac{c_{l} c_{r}}{(c_{l}-v)(v-c_{r})} \\
%\times
%\frac{k^2}
%{\cosh^2\left[\frac{k}{2}\left(\frac{x}{c_{l}-v}
%-\frac{x'}{v-c_{r}}\right)\right]}+O(x-x')^{-2}\,.
\end{multline}

\begin{figure}[t]
\begin{center}
\includegraphics[ width=70mm , height=50 mm]{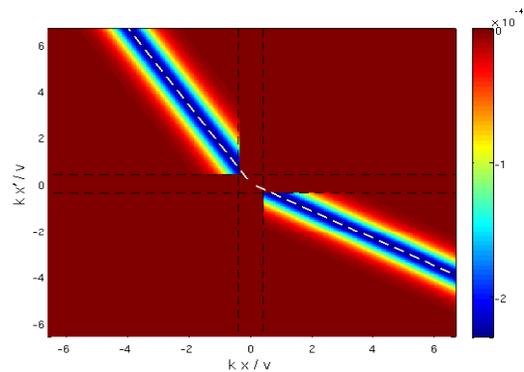}
\caption{Density-density correlation pattern
 $\left|G^{1D}_2(x,x')\right|/n_{1D}^2$. The valley-shaped feature
 indicated as a white dashed line is the signature of Hawking
 radiation. The black dashed lines indicate the $[-x_0,x_0]$ horizon
 region.}
\label{BH}
\end{center}
\end{figure}
The $x,x'$ dependence contained in the $\cosh$ term is the central
result of the present Letter: as shown in Fig. \ref{BH}, $G^{1D}_2(x,x')$ has
a quite narrow, valley-shaped feature centered on the
$x'/(v-c_l)=-x/(c_r-v)$ straight line that describes correlations
between pair of points located respectively inside and outside the black
hole. The bottom value of the valley goes as the squared surface gravity
and it is inversely proportional to the diluteness parameter $n_{1D}\xi$ of
the gas. The neglected subleading term $O[|x-x'|^{-2}]$ contains the
vacuum contribution of the ingoing modes and it is a consequence of the
short-range repulsion of bosons~\cite{castin}.
On the other hand, no specific signal appears for pair of points on the
same side of the horizon.

The position of the valley has a transparent physical interpretation: at
all times, pairs of Hawking phonons almost simultaneously emerge from the
horizon region and propagate into the sub- and super-sonic regions at
speeds respectively $c_r-v$ and $v-c_l$. After a propagation time
$\Delta t$, the initial quantum correlation then reflects into a density
correlation between $x=(c_r-v)\Delta t$ and $x'=-(v-c_l)\Delta t$.
The valley-shaped feature is immediately recovered once we integrate over
all possible values of $\Delta t$.
The width of the valley is proportional to the inverse of the surface
gravity $k^{-1}$ and it provides a simple way of estimating the Hawking
temperature of the emission. Note that under the hydrodynamic assumption the width is much larger than the healing length $\xi$. 

The contribution of mode back-scattering, here neglected,
quickly vanishes in the hydrodynamic limit: the corresponding
terms are in fact of higher order in $k$.
Although all our predictions have been obtained for the simplest
configuration, the behavior of the correlation function at late ($t\gg
k^{-1}$) times can be shown to be generic and independent on the details
of the horizon formation process \cite{c-QFT}.

We conclude by providing quantitative estimations of the Hawking signal
for parameters inspired to an existing guided Rb-atom laser
experiment~\cite{atomlaser}. 
For such systems, we can consider flow speeds on the order of
$v=4\,\textrm{mm/s}$ and sound speed modulations such that
$c_l/v=0.7$ and $c_r/v=1.5$, which implies that the healing length $\xi$
is on the order of $0.2\,\mu\textrm{m}$. A value
$2x_0=6\xi=1.2\,\mu\textrm{m}$ for the thickness of the modulation region
is expected to be enough to fulfill the hydrodynamic assumption.
Inserting these values into the expression for $T_H$, one obtains a
quite low~\cite{ketterle_picoK} Hawking temperature in the
$T_H\approx 4 \,\textrm{nK}$ range.
For a reasonable value of the diluteness parameter $n_{1D}\xi\simeq 10$, the maximum of
the density correlation signal (\ref{corr}) turns out to be
$G^{1D}_2/n_{1D}^2\sim -2.5 \times 10^{-4}$.
However, as our predictions follow from quantum hydrodynamics, their validity
goes beyond weakly interacting BECs and in particular in one dimension
they extend to a generic Luttinger liquid~\cite{density_phase}: stronger
correlation signals can therefore be observed in stronger interacting
systems where the diluteness parameter is smaller.

Furthermore, its peculiar valley shape should allow to isolate its
contribution on top of the background of thermal fluctuations~\cite{thermal} and
experimental noise. To this purpose several detection schemes have been
developed during recent years to experimentally characterize local
density fluctuations in ultracold atomic
gases~\cite{jin,HBTlattice,HBT_aspect}, some of which are able to
resolve even individual atoms~\cite{atom_detect_cavity}.
To overcome the atomic shot-noise, one may take advantage of the large
number of atoms that are available in the quasi-CW atom laser
beam~\cite{atomlaser,atomlaser2}.

In summary, we have shown that the gravitational analogy predicts a
characteristic peak in the density-density correlation function
in a flowing atomic Bose-Einstein condensate in the presence of a
horizon: the quantum correlation within the pairs of Hawking phonons
reflects into long-ranged density correlations that extend far
from the horizon. This constitutes a promising signature to detect and
isolate the Hawking emission of phonons from acoustic black holes.

We are grateful to R. Parentani and W. Unruh for stimulating
discussions. 
SF thanks ``E. Fermi'' Center for supporting her research. 
AR and IC thank financial support from EuroQUAM-FerMix.

{\bf Note added} During the review process, numerical
evidence supporting the predictions in this Letter has been obtained
starting from a microscopic and {\em ab initio} theory of the atomic
condensate~\cite{numbec}.

\end{document}